\documentclass[aps,prd,twocolumn,nofootinbib,superscriptaddress,preprintnumbers]{revtex4}
\usepackage[hypertex]{hyperref}
\usepackage{graphicx}
\usepackage{amsmath}
\usepackage{amssymb}
\usepackage[usenames]{color}

\newcommand{\bbar}[1]{{\,\overline{\!#1}{}}}
\newcommand{\uude}[4]{\bbar{U}_{#1} \bbar{U}_{#2} \bbar{D}_{#3}
\bbar{E}_{#4}\,}
\newcommand{\qqql}[4]{Q_{#1} Q_{#2} Q_{#3} L_{#4}\,}
\newcommand{\QQQL}{$Q_i Q_j Q_k L_l$\ }
\newcommand{\UUDE}{$\bbar{U}_i \bbar{U}_j \bbar{D}_k \bbar{E}_l$\ }
\newcommand{\orderone}{$\mathcal{O}(1)$\ }
\newcommand{\eps}{\epsilon}
\newcommand{\beq}{\begin{equation}}
\newcommand{\eeq}{\end{equation}}
\newcommand{\beqn}{\begin{eqnarray}}
\newcommand{\eeqn}{\end{eqnarray}}
\newcommand{\bsym}{\boldsymbol}
\newcommand{\mcal}{\mathcal}
\begin{document}

\pagestyle{plain}
\preprint{LBNL-54961}

\title{Probing the Planck Scale with Proton Decay\\~}

\author{Roni Harnik}
\affiliation{Theoretical Physics Group, 
Lawrence Berkeley National Laboratory, 
University of California, Berkeley, CA 94720, USA}
\affiliation{Department of Physics, University of California,
Berkeley, CA 94720, USA}

\author{~Daniel T. Larson}
\affiliation{Theoretical Physics Group, 
Lawrence Berkeley National Laboratory, 
University of California, Berkeley, CA 94720, USA}
\affiliation{Department of Physics, University of California,
Berkeley, CA 94720, USA}

\author{~Hitoshi Murayama}
\thanks{On leave of absence from Department of Physics,
University of California, Berkeley, CA 94720, USA.}
\affiliation{\mbox{Institute for Advanced Study, Princeton, 
NJ 08540, USA}\\[2mm]
\mbox{\textnormal{\texttt{
roni@socrates.berkeley.edu,
dtlarson@socrates.berkeley.edu,
murayama@ias.edu,
thor@th.physik.uni-bonn.de
}}}\\~}

\author{~Marc Thormeier}
\thanks{New address as of April 2004: Physikalisches Institut der
Universit\"at Bonn, Nu\ss allee 12, 53115 Bonn, Germany.}
\affiliation{Theoretical Physics Group, 
Lawrence Berkeley National Laboratory,
University of California, Berkeley, CA 94720, USA}
\affiliation{\mbox{Institute for Advanced Study, Princeton, 
NJ 08540, USA}\\[2mm]
\mbox{\textnormal{\texttt{
roni@socrates.berkeley.edu,
dtlarson@socrates.berkeley.edu,
murayama@ias.edu,
thor@th.physik.uni-bonn.de
}}}\\~}
~\\
\date{\today}

\begin{abstract}~\\\noindent
  We advocate the idea that proton decay may probe physics at the
  Planck scale instead of the GUT scale. This is possible because
  supersymmetric theories have dimension-5 operators that can induce
  proton decay at dangerous rates, even with $R$-parity
  conservation. These operators are expected to be suppressed by the
  same physics that explains the fermion masses and
  mixings. We present a thorough analysis of nucleon partial
  lifetimes in models with a string-inspired anomalous $U(1)_X$ family
  symmetry which is responsible for the fermionic mass spectrum as
  well as forbidding $R$-parity violating interactions. Protons and
  neutrons can decay via $R$-parity conserving non-renormalizable
  superpotential terms that are suppressed by the Planck scale and
  powers of the Cabibbo angle. Many of the models naturally lead to
  nucleon decay near present limits without any reference to grand
  unification.
\end{abstract} 
\maketitle

\section{Introduction}
\label{sec:intro}

\noindent
Baryon number is an accidental symmetry of the Standard Model (SM).
However, it is unlikely that baryon number will remain a symmetry up
to the highest energy scales.  Because of the gauge hierarchy problem
it is strongly believed that the SM must be augmented by new physics
at the TeV scale. There is no theoretical reason to believe that such
new physics will still conserve $U(1)_B$. In fact, baryon number
violation is one of the necessary ingredients in models that explain
the matter-antimatter asymmetry in the universe~\cite{Sakharov:dj}.
We therefore expect $U(1)_B$ to be violated at some higher energy
scale.  Current and future experiments of nucleon decay may thus be
probes of high-scale physics. For example, the partial lifetime of the
decay mode $p\to K^+ \bar{\nu}$ is greater than $1.6\times 10^{33}$
years~\cite{Kearns}, which places stringent constraints on scenarios
of new physics.

On the other hand, supersymmetry is considered the leading candidate
for physics beyond the SM because it solves the gauge hierarchy
problem once and for all and it is also consistent with grand unified
theories (GUTs).  It is often said that since quarks and leptons lie
in the same GUT multiplets, an observation of proton decay will be a
signal of a GUT.  Indeed, the minimal supersymmetric $SU(5)$ GUT has
been excluded because the unification of gauge couplings forces the
model into a region of parameter space where the proton decays too
quickly \cite{Murayama:2001ur}.

However, it is an under-publicized fact that supersymmetric models
predict proton decay \emph{even without unification}. Conventionally
$R$-parity ($R_p$) is imposed as an exact symmetry in order to prevent
the proton from decaying through renormalizable operators.
But in a generic supersymmetric model one can still write down
$R_p$-conserving, yet baryon and lepton number violating, $D=5$
operators suppressed by a single power of the cutoff scale, which we
take to be the reduced Planck scale~$M_\mathrm{Pl} \sim
2.4\times10^{18}$~GeV. These operators come from the superpotential
\begin{equation}
\label{eqn:w5}
W_5 = \frac{\epsilon_{abc}}{M_\mathrm{Pl}}\left( C^{ijkl}_L 
(Q_i^a Q_j^b)( Q_k^c
L_l) + C^{ijkl}_R  \bbar{U}_i^a \bbar{U}_j^b
\bbar{D}_k^c \bbar{E}_l \right).
\end{equation}
Here $Q,L$ are quark and lepton doublet superfields and $\bbar U,
\bbar D, \bbar E$ are $SU(2)_L$ singlets. The $SU(2)_L$ indices are
contracted between the terms in parentheses, while $i,j,k,l$ are
generational indices and $a,b,c$ are color indices.  From an effective
field theoretic point of view one expects these operators to appear
with coefficients $C_L$ and $C_R$ of $\mathcal{O}(1)$.

Supersymmetry's dirty little secret is that if one were to pick
generic $\mathcal{O}(1)$ numbers for the coefficients $C_L$ and $C_R$,
the proton lifetime would be about $10^{17}$ years, many orders of
magnitude below the experimental limit. Thus the coefficients $C_L$
and $C_R$ must be highly suppressed in any viable supersymmetric
model.

We do not consider this embarrassment a death-blow for supersymmetric
models, however. This is because the SM already contains highly
suppressed dimensionless numbers, namely the Yukawa coupling
constants. It is unsatisfactory to have an effective theory that is
valid up to the Planck scale without explaining the origin of small
coefficients for $H_\mathcal{U}Q_i\bbar U_j$, $H_\mathcal{D} Q_i\bbar
D_j$ and $H_\mathcal{D}L_i\bbar E_j$.  Such an explanation should
naturally determine the coefficients of $Q_iQ_jQ_kL_l$ and $\bbar
U_i\bbar U_j\bbar D_k\bbar E_l$, likely suppressing them as
well~\cite{Murayama:1994tc}. In the context of supersymmetric models
the most promising scenario for generating small Yukawa couplings is
that of Froggatt and Nielsen \cite{Froggatt:1978nt} (see
\cite{Dreiner:2003hw} for an exhaustive list of references), for which
the suppression of proton decay has been
demonstrated~\cite{Murayama:1994tc,Kakizaki:2002hs}. In
\cite{Dreiner:2003hw} the tight experimental bounds on exotic
processes were brought to bear on certain Froggatt-Nielsen models
found in the literature. Indeed, many models were found to be
incompatible with data solely due to insufficient suppression of
$Q_iQ_jQ_kL_l$. The relation between the fermion mass hierarchy in the
SM and the suppression of proton decay also occurs in other
non-supersymmetric models such
as~\cite{Arkani-Hamed:1999dc,Kakizaki:2001ue}.

In order to predict proton decay rates originating from Planck scale
$D=5$ operators we have to specify the framework of fermion masses and
mixings. We would like the framework to be specific enough so that we
can make quantitative predictions, but also general enough so that we
can study the generic features of Planck scale proton decay.  In this
paper we will focus on a recently proposed class of models based on a
single, anomalous Froggatt-Nielsen flavor symmetry
\cite{Dreiner:2003yr}.  In this class of models the MSSM superfields
are charged under a horizontal $U(1)_X$ symmetry that is spontaneously
broken by the non-zero VEV of a flavon field, $A$.  The MSSM Yukawa
terms are then suppressed by the ratio $\epsilon = \langle A \rangle
/M_\mathrm{Pl}$ raised to the the appropriate power necessary to
conserve $U(1)_X$.  The models in \cite{Dreiner:2003yr} are ambitious:
\begin{enumerate}
\item The $U(1)_X$ is the only symmetry beyond the SM gauge groups.
\item The only fields charged under the SM gauge groups are those in
the MSSM.  However, there may be (hidden sector) fields charged only
under $U(1)_X$.
\item The $U(1)_X$ charge assignments and breaking scale are inspired
directly from string theory.  Anomalies are canceled by the
Green-Schwarz mechanism which places sum rules on the $X$-charges.
\item The Cabibbo angle is \emph{calculated}. It is set by the flavon
VEV which is is determined by string theory, yielding the
phenomenologically interesting value~$\epsilon = \langle A \rangle
/M_\mathrm{Pl}= 0.17$--0.22.
\item The charge assignments are chosen to yield the measured quark
and lepton masses, including CKM mixing.
\item Neutrino mixings are also a result of the charge assignments.
\item $R$-parity is an exact (accidental) symmetry of the $U(1)_X$
charge assignments, preventing tree-level nucleon decay.
\end{enumerate}
That these seven goals can be achieved within a single, simple
framework is non-trivial and encouraging. As was shown
in~\cite{Dreiner:2003yr}, there is only a finite number of $X$-charge
assignments with these properties.\footnote{With further assumptions,
discussed in~\cite{Dreiner:2003yr}, these models can also explain the
scale of neutrino masses by invoking only two right-handed neutrinos
and using only two fundamental mass scales, $M_\mathrm{Pl}$ and
$m_\mathrm{soft}\sim 1$ TeV.} It is very interesting that the
predicted nucleon partial lifetimes in these models are generically
within a few orders of magnitude of the current limits of about
$10^{32}$ years. This already rules out some specific charge
assignments, and means that others will be directly tested very soon.
It is important to note that in general these models do \emph{not}
have GUT-compatible $X$-charges, but they nonetheless lead to very
interesting nucleon decay rates not much smaller than those predicted
by GUTs.

In this paper we systematically explore the nucleon partial lifetimes
predicted by the $U(1)_X$ Froggatt-Nielsen models presented
in~\cite{Dreiner:2003yr}. In Section~\ref{sec:fn} we briefly review
these models, then in Section~\ref{sec:estimates} we do some quick
estimates of the lifetimes. In Section~\ref{sec:decay} we describe our
method of computing the nucleon decay rates. Section~\ref{sec:results}
contains our main results, namely that the $U(1)_X$ models of flavor
are already constrained by current nucleon decay data, and they will
continue to be tested by future experiments. Our conclusions are
presented in Section~\ref{sec:conclusion}, while
Appendix~\ref{sec:dressing} carefully explains the dressing diagrams
relevant for our analysis, and Appendix~\ref{sec:kahler} discusses the
effects of canonicalizing the K\"ahler potential and transforming from
the interaction basis into the mass basis.

\section{Froggatt-Nielsen Framework}
\label{sec:fn}
\noindent
We will analyze the class of Froggatt-Nielsen models presented in
\cite{Dreiner:2003yr}. In this section we summarize the results that
are relevant for our analysis and refer readers to that paper for
details. Below the reduced Planck scale the gauge group and particle
content are those of the MSSM, with the addition of two right-handed
neutrinos, a single flavon superfield $A$, and a generation-dependent
$U(1)_X$ gauge group with anomalies canceled by the Green-Schwarz
mechanism~\cite{Green:sg}. Since all matter fields are charged under
$U(1)_X$, their couplings to the flavon are determined by $U(1)_X$
invariance. The goal is to have the flavon couplings generate the
generation dependent masses and mixings in the fermion spectrum.

For example, masses for the up-type quarks originate from the
non-renormalizable operator
\begin{equation}
  \label{huns}
  {g_{U}}^{ij}~
  \bigg(\frac{A}{M_\mathrm{Pl}}
  \bigg)^{{X_{H_\mathcal{U}}+X_{Q_i}+X_{\overline{U}_j}}} 
  H_\mathcal{U}\,Q_i\,\overline{U}_j,
\end{equation}
where the flavon charge is defined to be $X_A=-1$ and ${g_{U}}^{ij}$
are \orderone coefficients that are undetermined within the effective
theory below $M_\mathrm{Pl}$.  The dimensionless ${g_{U}}^{ij}$ is zero if
$X_{H_\mathcal{U}}+X_{Q_i}+X_{\overline{U}_j}$ is negative or
fractional.  The powers of $A$ in Eqn.~(\ref{huns}) compensate the
$U(1)_X$ charges of $ H_\mathcal{U}\, Q_i\,\overline{U}_j$ yielding
$U(1)_X$ gauge invariants.  The Dine-Seiberg-Wen-Witten-generated
Fayet-Iliopoulos term for $U(1)_X$ induces a VEV for the scalar
component of the flavon
field~\cite{Dine:bq,Dine:1986zy,Atick:1987gy,Dine:1987gj}. The VEV is
given by $\langle A \rangle = \epsilon M_\mathrm{Pl}$, where
$\epsilon$ naturally turns out to be the size of the sine of the
Cabibbo angle, $\eps\sim 0.2$. After $U(1)_X$-breaking one gets the
Yukawa couplings
\begin{equation}
  {G_{\!U}}^{ij}~=~{g_{U}}^{ij} ~
  \epsilon^{X_{H_\mathcal{U}}+X_{Q_i}+X_{\overline{U}_j}},
  \label{eff}
\end{equation}
which are suppressed by powers of $\eps$. In this way
the charge assignments of the MSSM matter fields determine the fermion
mass hierarchy. Once the $X$-charges are chosen to reproduce the mass
spectrum, the $\epsilon$-suppressions of other superpotential terms are
determined up to \orderone coefficients, with higher-dimensional
operators like $Q_iQ_jQ_kL_l$ being suppressed by additional powers of
$M_\mathrm{Pl}$. It is this connection that we will be exploiting to
estimate the nucleon decay lifetimes within this Froggatt-Nielsen
framework.

The $X$-charges are determined by requiring cancellation of chiral
anomalies between $U(1)_X$ and the SM gauge group, prohibition of all
$R_p$-violating interactions,\footnote{All $R_p$-even terms are
  supposed to have an overall integer $X$-charge, while all $R_p$-odd
  terms are supposed to have an overall half-odd-integer $X$-charge.}
and generation of the phenomenologically viable fermion masses and the
CKM matrix. The $X$-charges of the MSSM superfields are then specified
by six integer parameters, as demonstrated in~\cite{Dreiner:2003yr} and
displayed in their Table~2. These $X$-charges are further constrained
by the phenomenology of neutrino mixing, which fixes two of those six
parameters . The $X$-charge of each superfield is determined by the four
remaining parameters, $x, y, z$ and $\Delta^{\!H}$, and is shown in
Table~\ref{tab:xcharges}. The physical significance of these
parameters will be explained below.
\begin{table}
  \begin{center}
    \begin{tabular}{|rcl|}
      \hline
      & & \\
      $~~~\phantom{\Big|}X_{H_\mathcal{D}}$
      &$=$&$\frac{-24 + 12 y + z
        (9 + 4 z) + x (4 x + 22 + 6 z) -
        2  \Delta^{\!H}  (2 x + 12 + 3z)}{10~(6+x+z )  }$\\
      $~\phantom{\Big|}X_{H_\mathcal{U}}$
      &$=$&$-z-\phantom{\Big|}X_{H_\mathcal{D}}  $\\
      $~\phantom{\Big|}X_{Q^1}$&$=$&$\frac{1}{3}\Big(
      \frac{19}{2} - X_{H_\mathcal{D}} + x + 2y + 2z -
      \Delta^{\!H}\Big)$\\
      $~\phantom{\Big|}X_{Q_2}$&$=$&$X_{Q^1}-1-y $\\
      $~\phantom{\Big|}X_{Q_3}$&$=$&$X_{Q^1}-3-y $\\
      $~\phantom{\Big|}X_{\overline{D}_1}$&$=$&$-X_{
        H_\mathcal{D}}-X_{Q_1}+4+x $\\
      $~\phantom{\Big|}X_{\overline{D}_2}$&$=$&$X_{\overline{D}_3}=X_{
        \overline{D}_1}-1+y $\\
      $~\phantom{\Big|}X_{\overline{U}_1}$&$=$&$
      X_{H^\mathcal{D}}-X_{Q_1}+8+z $\\
      $~\phantom{\Big|}X_{\overline{U}_2}$&$=$&$
      X_{\overline{U}_1}-3+y $\\
      $~\phantom{\Big|}X_{\overline{U}_3}$&$=$&$
      X_{\overline{U}_1}-5+y $\\
      $~\phantom{\Big|}X_{L_1}$&$=$&$\frac{1}{2}+
      X_{H_\mathcal{D}}+\Delta^{\!H}$\\
      $~\phantom{\Big|}X_{L_2}$&$=$&$X_{L_3} = X_{L_1} - z$\\
      $~\phantom{\Big|}X_{\overline{E}_{1}}$&$=$&$-X_{
        H_\mathcal{D}}        {+~4} - X_{L_1} + x + z $\\
      $~\phantom{\Big|}X_{\overline{E}_2}$&$=$&$-X_{
        H_\mathcal{D}}  {+~2} - X_{L_1} + x+ z      $ \\
      $~\phantom{\Big|}X_{\overline{E}_{3}}$&$=$&$-
      X_{H_\mathcal{D}}  \phantom{+~4} - X_{L_1} +
      x+z            $  \\
      & &\\ \hline
    \end{tabular}
    \caption{$R_p$-conserving $X$-charge assignments of the MSSM
      superfields in terms of four parameters: $x$, $y$, $z$, $\Delta^H$. The
      charge assignments are as in Table~2 of~\cite{Dreiner:2003yr} with their 
      $\zeta$ and $\Delta^L_{31}$ both taken to be $-z$ as required by
      neutrino mixing phenomenology. }
    \label{tab:xcharges}
  \end{center}
\end{table}

Only the three parameters $x$, $y$ and $z$ are relevant for nucleon
decay.  They are restricted to a small set of integers, which leads to
24 distinct models with different nucleon decay signatures. The
allowed values are $x=0$,~$1$,~$2$,~$3$; $y=-1$,~$0$,~$1$,
\footnote{In Ref.~\cite{Dreiner:2003hw} $y=-7,-6$ are also considered
because at first sight they give a viable CKM matrix once the
supersymmetric zeros are filled in. However,
Ref.~\cite{Espinosa:2004ya} demonstrates that this is not the
case. See also Refs.~\cite{King:2004tx}. Therefore we do not consider
$y=-7,-6$ in this paper.}
and $z=0$,~$1$.

The parameter $x$ is related to the ratio of the bottom and top quark masses,
and is thus also connected to $\tan\beta$,
\begin{equation} 
  \frac{m_b}{m_t} \sim \frac{\eps^x}{\tan\beta}. 
\end{equation} 
Since $\eps\sim 0.2$, larger $x$ corresponds to smaller $\tan\beta$,
but because of unknown \orderone coefficients, we cannot determine
$\tan\beta$ exactly. To simplify our analysis we will choose a
specific, reasonable value of $\tan\beta$ for each value of $x$,
namely $\tan\beta=50$, $20$, $5$, $3$ for $x=0, 1, 2, 3$,
respectively.

The three choices for the parameter $y$ determine the texture of the
up- and down-quark Yukawa matrices and therefore also the CKM matrix.
One finds
\begin{eqnarray}
  y=-1:~ && \boldsymbol{V_{\!C\!K\!M}} \sim \left(\begin{array}{ccc} 1 & 1 & 
      \eps^2 \\ 1 & 1
      & \eps^2 \\ \eps^2 & \eps^2 & 1 \end{array}\right), \\
  y=0:~ && \boldsymbol{V_{\!C\!K\!M}} \sim \left(\begin{array}{ccc} 1 & 
      \eps & \eps^3 \\
      \eps & 1 & \eps^2 \\ \eps^3 & \eps^2 & 1 \end{array}\right), \\
  y=1:~ && \boldsymbol{V_{\!C\!K\!M}} \sim \left(\begin{array}{ccc} 1 & 
      \eps^2 &
      \eps^4 \\
      \eps^2 & 1 & \eps^2 \\ \eps^4 & \eps^2 & 1 \end{array}\right).
\end{eqnarray}

Finally, the parameter $z$ is related to the charged lepton mass
spectrum,
\begin{equation}
  \frac{m_e}{m_\mu}\sim\eps^{2+z}.
\end{equation}
In combination with neutrino phenomenology and the requirement that
$R_p$ is conserved by virtue of the $X$-charges, it turns out that $z$
specifies the texture of the MNS matrix:
\begin{equation}
  \boldsymbol{U_{\!M\!N\!S}} \sim \left(\begin{array}{ccc} 1 & \eps^z & \eps^z 
      \\ \eps^z & 1
      & 1 \\ \eps^z & 1 & 1 \end{array}\right).
\end{equation}
Here we see that $z=0$ corresponds to an ``anarchical'' MNS
matrix~\cite{Hall:1999sn,Haba:2000be}, whereas $z=1$ corresponds to a
``semi-anarchical'' MNS matrix \cite{Sato:1997hv}.

The minimalist approach is to add only two right-handed neutrinos.
When this approach is taken the fourth parameter, $\Delta^{\!H}$, can
take two different values for each set of $\{x,y,z\}$, but it
has no impact on nucleon lifetimes and so it is irrelevant for our
purposes. However, our analysis does not depend on the number of right handed
neutrinos.

In addition to specifying the CKM and MNS textures, the parameters
$\{x,y,z\}$ also have bearing on other aspects of the UV physics. For
example, $SU(5)$ invariance is only consistent with $y=1$ and $z=0$.
Also, $z=1$ prohibits the operator $H_\mathcal{D} H_\mathcal{U}$, thus
allowing the $\mu$-term to be generated by the Giudice-Masiero
mechanism \cite{Giudice:1988yz,Kim:1994eu}. If one has a preference
towards a specific physical scenario, the number of models to choose
from is significantly reduced.  For example, the models
in~\cite{Dreiner:2003yr} were chosen to have $z=1$ for the sake of a
natural $\mu$-parameter, $y=0$ to obtain the most natural CKM matrix,
and $x=2,3$ to avoid proton decay limits.

Other criteria may be used in choosing one's favorite model. For
instance, the values of the $X$-charges are much more aesthetically
pleasing in some models than others. For example, taking $x=2$, $y=1$, $z=0$
and $\Delta^{\!H}=9$, one finds the $X$-charges of all MSSM
superfields are integers or half-odd-integers, as shown in
Table~\ref{TableII}.  However, in
this paper we will treat all 24 distinct models equally and focus
only on their predictions for nucleon decay.
\begin{table}[ht]
  \begin{tabular}{|c|c|c|c|}
    \hline
    $\phantom{\Big|}$Generation $i\phantom{\Big|}$ & 1 & 2 & 3 \\ \hline
    $\phantom{\Big|}X_{Q_i}\phantom{\Big|}$ & $5/2$ & $1/2$ & $-3/2$ \\
    $\phantom{\Big|}X_{L_i}\phantom{\Big|}$ & $13/2$ & $13/2$ & $13/2$ \\
    $\phantom{\Big|}X_{\bbar{U}_i}\phantom{\Big|}$ & $5/2$ & $1/2$ & 
    $-3/2$ \\
    $\phantom{\Big|}X_{\bbar{D}_i}\phantom{\Big|}$ & $13/2$ & $13/2$ 
    & $13/2$ \\
    $\phantom{\Big|}X_{\bbar{E}_i}\phantom{\Big|}$ & $5/2$ & $1/2$ & 
    $-3/2$ \\ 
    \hline
  \end{tabular}
~\\
\begin{tabular}{|c|}\hline
  $\phantom{\Big|}X_{H_\mathcal{D}}=-3$, $X_{H_\mathcal{U}}=
  3\phantom{\Big|}$\\
  \hline\end{tabular}
\caption{\label{TableII}
  An example of an $X$-charge assignment for the matter superfields in
  the model with
  $x=2$, $y=1$, $z=0$, and $\Delta^{\!H}=9$. $\Delta^{\!H}$ helps to
  determine the specific charges, but its effects cancel out in all
  the nucleon decay operators.}
\label{tab:charges}
\end{table}

\section{Quick Estimates}
\label{sec:estimates}

A quick estimate shows that the operator $Q_1 Q_1 Q_2 L_3$ leads a
proton lifetime of $\sim 10^{17} \eps^{-2n}$ years, where $n=9+x+y+z$
is the sum of the $X$-charges for the four superfields. Since
$\eps\sim 0.2$, we need to have $n\gtrsim 11$ to satisfy the
experimental limit of about $10^{33}$ years for $p\to
K^+\bar{\nu}_{\tau}$. This demonstrates the general trend that larger
values of $x$, $y$, and $z$ will lead to greater suppression of the
decay rate and hence a longer lifetime. 

It turns out that the nucleon lifetimes predicted by our
Froggatt-Nielsen model are in the same ballpark as those predicted by
GUT models. This can be understood relatively easily. For the operator
$Q_2 Q_2 Q_1 L_2$ we can rewrite the suppression in terms of Yukawa
couplings (ignoring \orderone factors),
\begin{equation}
  \frac{1}{M_\mathrm{Pl}} \eps^{8+x+z} = \frac{\eps}{M_\mathrm{Pl}}
  \eps^z \eps^{4} \eps^{2+x} \eps = \frac{\eps}{M_\mathrm{Pl}}
  \frac{\mu}{m_{3/2}}  {G_{\!U}}^{22}~{G_{\!E}}^{22} \eps,
\end{equation}
where we have used the phenomenological constraints on the $X$-charges
from Section~5
in~\cite{Dreiner:2003yr} to identify the Yukawa couplings. This is to
be compared with the coefficient of the same operator coming from an
$SU(5)$ GUT theory, which is
\begin{equation}
  \frac{1}{M_\mathrm{GUT}} {G_{\!U}}^{22}~{G_{\!E}}^{22}  \lambda_C .
\end{equation}
Equating $\eps$ with $\lambda_C$ and noting that the $\mu$-parameter
cannot be too different from the gravitino mass, the difference
between the coefficients is simply $\lambda_C
(M_\mathrm{GUT}/M_\mathrm{Pl})$. Considering the fact that these two models
involve completely different physics, we find this difference
interestingly small.  The estimation above is obviously very crude and
calls for a more rigorous analysis which is the goal of this work.

\section{Nucleon Decay Computations}
\label{sec:decay}

\begin{figure}[tc]
  \centering \includegraphics[width=.8\columnwidth]{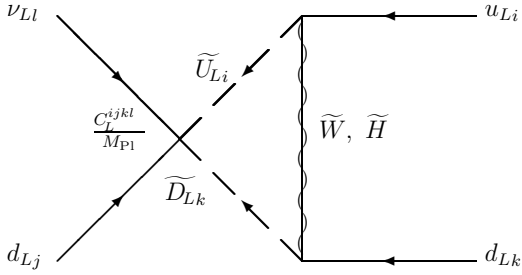}
  \caption{An example of a dimension-5 operator that can be dressed by
    neutral or charged winos and higgsinos.}
  \label{fig:dress}
\end{figure}

\noindent
In this section we describe our calculations of the various nucleon
decay modes. We use the standard methods for the computations,
following~\cite{Murayama:1994tc} and~\cite{Goto:1998qg}. The starting
point is the higher-dimensional UV Lagrangian shown in
Eqn.~(\ref{eqn:w5}).  The dimension-five fermion-fermion-scalar-scalar
terms arising from these operators can be ``dressed'', whereby the
scalars exchange a gaugino or higgsino and are converted into
fermions, as shown in Figure~\ref{fig:dress}. Below the scale of the
soft masses they become four-fermion operators and contribute to
nucleon decay. The matrix elements of these operators can be evaluated
using the well-known chiral Lagrangian technique
\cite{Claudson:1981gh}.

The quark doublet superfields $Q$ in Eqn.~(\ref{eqn:w5})
are taken to be in the SuperCKM basis, namely
$Q=(U, D^\prime) = (U, V_{\!C\!K\!M} D)$ where
$U$ and $D$ are the mass eigenstates. Thus
the couplings to the mass eigenstate quark operators will come with
various CKM factors. However, we cannot necessarily neglect operators
with small CKM factors because they may be offset by the lesser degree of
$\eps$-suppression in $C^{ijkl}_{L,R}$. In other words, operators
$Q_iQ_jQ_kL_l$ containing third generation quarks will have small CKM
couplings to the first generation quarks in the nucleon, but will have
correspondingly larger coefficients $C^{ijkl}_{L,R}$ due to the
$X$-charge assignments for the third generation.

Note that the $U(1)_X$ basis is not necessarily the same as the
SuperCKM basis nor the mass eigenbasis. Thus it is important to take
these various changes of basis into consideration when determining the
$\eps$-suppression of the nucleon decay operators. For the most part
this only effects the \orderone coefficients, which we are ignorant of
anyway. Nevertheless, we have performed a thorough analysis of this
effect as described in Appendix~\ref{sec:kahler}.

The operators \QQQL and \UUDE are not all independent. Fierz
identities may be used to convert all of the operators into a smaller
set which we choose as our basis of independent operators. One can
further reduce the number of operators in this set by noting that
contributions to nucleon decay can only come from operators with at
least one first-generation superfield. The Fierz identities and a list
of independent operators are given in
Appendix~\ref{sec:dressing}. Since we do not know the exact
coefficients in front of these operators, we will assume that in the
UV all of the operators in our basis are generated and that their
unknown coefficients are given by the $\eps$-suppression determined
by the $X$-charges and \orderone numerical factors.

As argued above, the $\eps$-suppression is essential for the
predicted nucleon lifetimes to be above the current limits. This
suppression is determined by the $X$-charges of $Q_i$, $\bbar{U}_i$,
$\bbar{D}_i$, $L_i$ and $\bbar{E}_i$, which are in turn determined by
the choice of the model parameters $x$, $y$, and $z$.  Note that
$\eps$ itself also has mild dependence on the model parameters,
\begin{equation} 
  \epsilon ~=~\frac{g_3}{4\pi\sqrt{2}} \sqrt{3(6+x+z)}\,
\end{equation} 
where the QCD gauge coupling $g_3$ is roughly $0.72$ at high energies.
The dependence of the $\eps$-suppression on the choice of
$X$-charges leads to the different patterns of nucleon partial
lifetimes that we study in the next section.

Renormalization group effects enhance the nucleon decay operators due
to running between $M_\mathrm{Pl}$ and $m_p$.  For simplicity, we
compute the supersymmetric running from $M_\mathrm{Pl}$ down to $m_Z$
from the gauge couplings, ignoring corrections proportional to Yukawa
couplings. The left- and right-handed operators, $Q_iQ_jQ_kL_l$ and
$\bbar{U}_i\bbar{U}_j\bbar{D}_k\bbar{E}_l$, are enhanced by a factor
\begin{equation}
  \left(\frac{\alpha_1(M_\mathrm{Pl})}{\alpha_1(m_Z)}\right)^{c_1}
  \left(\frac{\alpha_2(M_\mathrm{Pl})}{\alpha_2(m_Z)}\right)^{c_2}
  \left(\frac{\alpha_3(M_\mathrm{Pl})}{\alpha_3(m_Z)}\right)^{c_3}.
\end{equation}
Here ${\alpha_i}^{-1}(q)={\alpha_i}^{-1}(m_Z)-\frac{b_i}{2\pi}\ln
\left(q/m_Z\right)$ with $b_i=(33/5,1,-3)$ and $c_i=(1/33, 3,
-4/3)$ for LH and $(2/11, 0, -4/3)$ for RH operators. Also, $g_1^2=(5/3)g'^2$
Numerically this enhances the LH operators by a factor of $9.3$ and
the RH operators by a factor of $5.6$.

In general there are many different ways the dimension-five operators
can be dressed to give four-fermion operators. In our scenario we
assume that supersymmetry breaking is sufficiently flavor blind so
that the flavor mixing in gluino-quark-squark vertices is negligible.
Then in the limit of degenerate squarks and degenerate sleptons the
contributions from gluino dressing cancel due to a Fierz identity, as
explained in more detail in Appendix~\ref{sec:dressing}. There it is
also shown that bino dressing does not contribute to nucleon decay.
All contributions from neutral-higgsino dressing are extremely
suppressed by small, first-generation Yukawa couplings, so we ignore
them. We do include the leading contributions from charged-higgsino
dressing. In the case of charged-lepton decay modes these
contributions turn out to be negligible, but~\cite{Goto:1998qg} has
shown that they can be important for the antineutrino decay modes. The
bulk of the contributions come from charged-wino dressing, though the
neutral-wino dressing can be comparable for the antineutrino decay
modes.

The dressing leads to a four-fermion vertex which must then be run
from the scale of the soft masses, roughly $10^2$~GeV, down to $1$~GeV
where they mediate nucleon decay. For simplicity we compute the
QCD running only from $m_Z$ down to $1$~GeV, which leads to an enhancement
given in~\cite{Buras:1977yy}, namely
\begin{equation}
  \left(\frac{\alpha_3(M_\mathrm{low})}{\alpha_3(M_\mathrm{high})}
  \right)^{2/\beta_0}
\end{equation}
where $\beta_0=11-\frac23 f$ and $f$ is the number of quark flavors
with masses less than $M_\mathrm{high}$. There are three of these
enhancement factors for the successive steps between $m_Z$, $m_b$,
$m_c$, and $m_p$. In
each step the QCD coupling is given by ${\alpha_3}^{(f)}(q) =
4\pi/(\beta_0 \ln(q^2/{\Lambda_f}^2))$, where $\Lambda_f$ is
determined by requiring that $\alpha_3$ be continuous between
successive intervals. Numerically we use $m_b=4.2$ GeV, $m_c=1.2$ GeV,
and $m_p=1$ GeV, which gives a total contribution of
$1.3$, and is quite insensitive to the exact thresholds for $m_b$ and
$m_c$. Together with the high scale running mentioned above, there is
a net enhancement of approximately $12$ for $C_L$ and $7$ for
$C_R$.

The chiral Lagrangian can be used to determine the decay rate induced
by the four-fermion operators~\cite{Claudson:1981gh}. The partial
decay width of a nucleon $B_i$ into a meson $M_j$ and lepton $l_k$ is
given by\footnote{Eqn.~(\ref{eqn:decaywidth}) is written in the
  notation of~\cite{Goto:1998qg}: $k$ is a generational index, but $j$
  refers to $\pi^0$, $\eta^0$, $K^0$, $\pi^\pm$ or $K^+$, and $i$
refers to
  proton or neutron.}
\begin{equation} \label{eqn:decaywidth}
  \Gamma(B_i \rightarrow M_j l_k) = \frac{m_i}{32\pi ~f_\pi^2}\left(
    1-\frac{{m_j}^2}{{m_i}^2} \right)^{\!2} \!\left(
    \big|A^{ijk}_L\big|^2 + \big|A^{ijk}_R\big|^2 \right)
\end{equation}
where $A^{ijk}_{L,R}$ can be found in Table~1 of~\cite{Goto:1998qg}, given in
terms of the coefficients of the various four-fermion operators
generated by dressing of the superpotential operators shown in
Eqn.~(\ref{eqn:w5}). The numerical values of $A_{L,R}$
depend on several input parameters. The coupling between baryons and
mesons in the chiral Lagrangian are given by $D$ and $F$. As
in~\cite{Claudson:1981gh} we take $D=0.81$, $F=0.44$, and $f_\pi=139$
MeV.

Our computations of nucleon lifetimes are subject to two types of
uncertainties, those that reflect our current lack of knowledge and
those that are inherent to the effective theory framework of our
model. The first type includes the uncertainty in the hadronic matrix
element and the masses of the superpartners, both of which we assume
will be pinned down more precisely by the time the next generation
nucleon decay experiments start taking data. The second type includes
the unknown \orderone coefficients that come with the
non-renormalizable superpotential operators, which cannot be
determined without a full theory of the Planck scale physics. We will
discuss each of these uncertainties in turn.

The value of the hadronic matrix element must be evaluated to convert
the four-fermion operators into nucleon lifetimes. In the framework of
the chiral Lagrangian this appears as two parameters $\beta_p$ and
$\alpha_p$, which are related by $\alpha_p=-\beta_p$. Unfortunately,
the value of $\beta_p$ is only known to roughly a factor of 10. The
conservative range often taken is $\beta_p = 0.003$--$0.03$ GeV$^3$.
Recently there has been some progress evaluating $\beta_p$ on the
lattice, yielding the results $\beta_p = 0.014\pm0.001$ GeV$^3$
in~\cite{Kuramashi:2000hw} and $\beta_p = 0.007\pm0.001$ GeV$^3$
in~\cite{Aoki:2002ji}. Note that the errors are only statistical, and
do not reflect systematic effects such as quenching. Since $\beta_p$
appears in the amplitude, any uncertainty in its value gets squared in
the decay rate or lifetime. However, this uncertainty is common across
all modes and all different models, so it simply represents an overall
rescaling of the lifetimes. For our computations we will use an
intermediate value, $\beta_p=0.01$~GeV$^3$. A smaller value of
$\beta_p$ would give a smaller decay rate and a \emph{longer}
lifetime. Allowing for the range $0.003$--$0.03$ thus corresponds to a
change in our results by a factor of 10 in both directions. Again, we
can look forward to reduction in the uncertainty in the hadronic
matrix element, which will reduce the uncertainty in our computations.

The scale of the superpartner masses is another uncertainty in our
computations that will be reduced if supersymmetry is discovered at a
collider.  The superpartner masses enter through the loop in the
dressed diagrams such as the one in Figure~\ref{fig:dress}. We assume
that the squarks and sleptons all have a common mass,
$m_\mathrm{soft}$. If the gaugino (or higgsino) mass is much less than
the squark mass, the contribution from the loop is given by
$M_\mathrm{gaugino}/m_\mathrm{soft}^2$, whereas if the gauginos are
degenerate with the squarks at $m_\mathrm{soft}$, then the loop factor
is $1/(2m_\mathrm{soft})$. One can imagine extremes where all the
superpartners are relatively light, near $100$~GeV, leading to the shortest
lifetimes, or the opposite extreme where the scalars are heavy (TeV) and the
gauginos are light (100~GeV), leading to the longest lifetimes.  

We can cover the different possibilities by taking all superpartners
to be degenerate at $m_{\mathrm{soft}}$ and then allowing
$m_\mathrm{soft}$ to range from 100 GeV to 10 TeV. The latter case
does not mean the superpartner masses are actually 10 TeV, but rather
represents the case where squark and slepton masses are around 1 TeV
and wino masses are 100 GeV.  In our computations we will choose a
middle ground by assuming of $m_\mathrm{soft}=1$~TeV.  With this
assumption, the proton lifetime simply scales inversely with
$m_\mathrm{soft}^2$. Choosing one of the other two scenarios described
above would change the nucleon lifetimes by a factor of 100.  We
reiterate that this uncertainty will disappear once the superpartner
spectrum is measured, most likely at the LHC.

Our ignorance of the \orderone coefficients appearing in front of each
operator in Eqn.~(\ref{eqn:w5}) is an uncertainty inherent in the
framework of effective field theory.  $A_{L,R}$ in
Eqn.~(\ref{eqn:decaywidth}) are each the sum of several contributing
amplitudes from different operators that can interfere with one
another. However, since we do not know either the exact magnitudes or
phases of the \orderone coefficients in the UV Lagrangian, we cannot
know whether the different contributions will interfere constructively
or destructively. As an extreme example, one could pick a fine-tuned
set of numbers such that the amplitude for a certain decay mode
vanishes completely. However, this is unlikely to occur in the real
world unless there is some symmetry of the high-energy theory that
enforces such a cancellation.

To remove the effects of such unlikely cancellations from our
computations we will add the various terms contributing to
$A_{L,R}$ in quadrature. This effectively gives the
average of many computations with different random phases between the
individual contributions. We take this as the central value of our
results. However, it is possible that even without fine-tuning there
could be significant effects due to interference between amplitudes.
In order to estimate this effect we perform the calculation in two
other ways, plotting the results as upper and lower error bars around
the central value. The lower error bar is determined by choosing all
the amplitudes to have the same phase, thereby interfering
constructively. The upper error bar is determined by choosing the
\orderone numbers to all be $+1$ or $-1$ such that amplitudes of
similar sizes interfere destructively. Even though these are not rigid
upper and lower bounds, they give a good sense of the possible
variations due to interference between amplitudes. However, it is
important to note that these are in no way $1\sigma$ error bars.

Finally, we will compare all of our computations of the partial
lifetimes to the experimental limits, which are taken
from~\cite{Kearns} and~\cite{Hagiwara:fs} and are shown in
Table~\ref{tab:modes}.
\begin{table}[ht!]
  \begin{tabular}{|l|c|}
    \hline
     & ~Expt. Limit ~ \\
     \raisebox{2.0ex}{~~~Mode} & \raisebox{1.0ex}{($10^{32}$ years)} \\ \hline
    $~\phantom{\Big|}p\rightarrow K^+ \bar{\nu}$ & $16$ \\
    $~\phantom{\Big|}p\rightarrow \pi^0 \mu^+$ & $37$ \\
    $~\phantom{\Big|}p\rightarrow K^0 \mu^+$ & $10$ \\
    $~\phantom{\Big|}p\rightarrow \eta^0 \mu^+$ & $7.8$ \\
    $~\phantom{\Big|}p\rightarrow \pi^0 e^+$ & $50$ \\
    $~\phantom{\Big|}p\rightarrow \pi^+ \bar{\nu}$ & $0.25$ \\
    $~\phantom{\Big|}p\rightarrow K^0 e^+$ & $5.4$ \\
    $~\phantom{\Big|}p\rightarrow \eta^0 e^+$ & $11$ \\
    $~\phantom{\Big|}n\rightarrow K^0 \bar{\nu}$ & $3.0$ \\
    $~\phantom{\Big|}n\rightarrow \eta^0 \bar{\nu}$ & $5.6$ \\
    $~\phantom{\Big|}n\rightarrow \pi^0 \bar{\nu}$ & $1.1$ \\
    $~\phantom{\Big|}n\rightarrow \pi^- \mu^+$ & $1.0$ \\
    $~\phantom{\Big|}n\rightarrow \pi^- e^+$ & $1.6$ \\ \hline
  \end{tabular}
  \caption{The various nucleon decay modes we consider, together with
    the experimental upper bound on the partial lifetime. Experimental
    values are from~\cite{Kearns} and~\cite{Hagiwara:fs}.}
  \label{tab:modes}
\end{table}

\section{Results}
\label{sec:results}

\noindent
In this section we present the results of our analysis of the nucleon
partial lifetimes within the context of the $U(1)_X$ flavor model of
\cite{Dreiner:2003yr}.  When the various amplitudes are added
incoherently, as described in the previous section, the most
constraining mode for all models is $p\rightarrow K^+ \bar{\nu}$,
because it has both a relatively large rate and a stringent
experimental bound.  First we will focus on this mode and then come
back to the more promising models and look at other decay modes in
more detail to see what we can expect to learn from nucleon decay
experiments in the near future.

As described in Appendix~\ref{sec:kahler}, the Froggatt-Nielsen
suppression of the operators $Q_iQ_jQ_kL_l$ and
$\bbar{U}_i\bbar{U}_j\bbar{D}_k\bbar{E}_l$ are not trivially
determined by $U(1)_X$ conservation, since they are affected by the
canonicalization procedure of the K\"ahler potential and the
transformation from the interaction basis into the mass basis.  In
Appendix~\ref{sec:kahler} we will show that this is not an issue.

\begin{figure}[tc]
  \centering \includegraphics[width=\columnwidth]{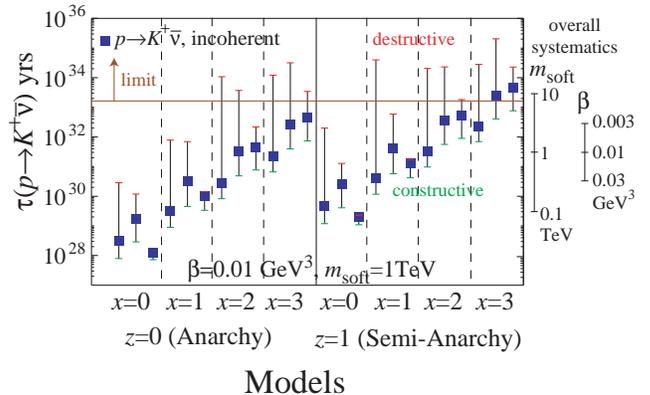} 
  \caption{Plot of proton partial lifetime in years for the mode
  $p\rightarrow K^+\bar{\nu}$, shown for models with $y=-1,0,1$.
  Models with $z=0\ (1)$ are shown on the left (right), and within
  each half $x$ increases from left to right, corresponding to a
  decrease in $\tan\beta$.  The error bars are \emph{not} $1\sigma$
  bars. Rather, the central point corresponds to incoherent addition of
  amplitudes. The upper and lower ends of the `error bar' are the
  lifetimes computed by adding the various amplitudes destructively
  and constructively, respectively. The horizontal line shows the
  experimental lower limit of $1.6\times 10^{33}$ years.  The
  calculation was done with $m_\mathrm{soft}=1$ TeV and
  $\beta_p=0.01$, but the scales on the right show the overall shift
  that would result from changing either $m_\mathrm{soft}$ or
  $\beta_p$.}  \label{fig:Knu-uncertainties}
\end{figure}

In Figure~\ref{fig:Knu-uncertainties} we show the partial lifetimes in
the $p\rightarrow K^+ \bar{\nu}$ mode for the 24 models,
using the three different methods of adding the different
contributions: incoherent (central value), constructive and destructive
(tips of error bars). The plot is divided in half, with $z=0$
on the left half and $z=1$ on the right half, and in each half the
value of $x$ increases from left to right, corresponding to a decrease
in $\tan\beta$. The current experimental bound is plotted as a
horizontal line. We have included scale bars at the right to indicate
the size of the overall systematic uncertainty due to
$m_\mathrm{soft}$ and $\beta_p$.

The first thing to notice is that many of the models are disfavored,
unless they are fine-tuned to give a highly destructive interference
or the supersymmetric spectrum is carefully chosen.  The models that
survive best are those with a high $x$ which corresponds to a lower
$\tan\beta$. In fact, of the models with $x=0$ or $x=1$, only one
model has a prediction above the current experimental limit even when
the amplitudes are added destructively.  Recall, however, that the
uncertainties in $\beta_p$ and $m_\mathrm{soft}$ are not included in
the error bar, which can potentially change the prediction by the
factors shown graphically to the right of Figure
\ref{fig:Knu-uncertainties}.  Nevertheless, it seems clear that models
with $x=0$ or $x=1$ are disfavored by current proton decay limits,
barring extremely delicate accidental cancellations, as was
anticipated in \cite{Dreiner:2003yr} based on a rough estimate.  For
this reason we will focus on models with $x=2$ or $x=3$ in what
follows. From Figure~\ref{fig:Knu-uncertainties} we also see that a
long proton lifetime favors a \emph{semi}-anarchical MNS matrix
($z=1$) and not one determined by anarchy ($z=0$). Note that the
preference for $z=1$ is also consistent with the Giudice-Masiero
mechanism that naturally produces a $\mu$-term of the right
size. However, this preference is only mild (within the uncertainty).

\begin{figure}[tc]
  \centering
  \includegraphics[width=\columnwidth]{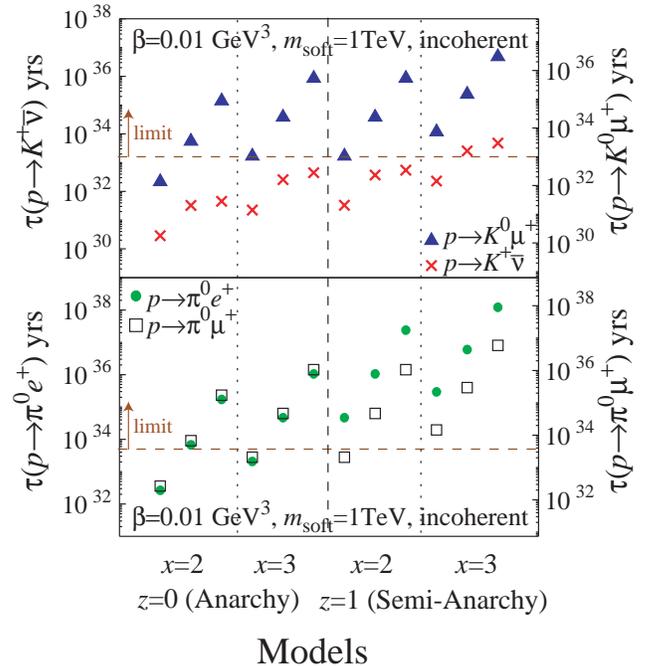}
  \caption{Comparison of proton lifetime in years for four different
    decay modes. The upper plot shows the computed lifetime for $p\to
    K^+ \bar{\nu}$ ({\color{Red}$\times$}, left axis) and $p\to K^0
    \mu^+$ ({\color{Blue}$\blacktriangle$}, right axis). The lower
    plot shows $p\to \pi^0 e^+$ ({\color{Green}$\bullet$}, left axis)
    and $p\to \pi^0 \mu^+$ ($\square$, right axis). For each plot the
    data for the two modes have been scaled so that the experimental
    limits, represented by the horizontal line, coincide, hence the
    different scales on the left and right axes.  The lifetimes are
    plotted for all twelve models with $\tan\beta \lesssim 10$.}
\label{fig:4modes}
\end{figure}

The current and upcoming experiments are expected to be able to detect
charged leptons with a high efficiency. It is thus interesting to also
focus on modes with charged leptons in the final state. In fact, for
the proton the next modes to appear after $p\to K^+ \bar{\nu}$ are
generally $p\to \pi^0 e^+$, $p\to \pi^0 \mu^+$, and $p\to K^0 \mu^+$.
In Figure~\ref{fig:4modes} we specialize to models with $x=2$ or $3$
(i.~e. $\tan\beta \lesssim 10$) and show the expected lifetime from
all four of the above decay modes. Here we plot the lifetime computed
using incoherent addition of amplitudes, but the uncertainty due to
unknown \orderone numbers should be kept in mind.  We see that most
models which survive the constraint from $p\to K^+ \bar{\nu}$ have a
lifetime for $p\to \pi^0\mu^+$ that is within two or three orders of
magnitude of the experimental bound, while $p\to K^0 \mu^+$ is only
slightly larger, and $p\to \pi^0 e^+$ can potentially be smaller. This
raises the exciting possibility of two or three decay modes being
detected in the coming round of experiments.  Figure~\ref{fig:4modes}
also shows how the relative decay rates can be used to distinguish
models with different phenomenology. For instance, all models with
$z=0$ (anarchy) have $p\to \pi^0 e^+$ as the dominant decay mode with
a charged lepton in the final state, whereas its lifetime is
substantially higher for the models with $z=1$. In both cases $p\to
\pi^0 \mu^+$ and $p\to K^0 \mu^+$ are comparable.

\begin{figure}[tc]
  \centering \includegraphics[width=\columnwidth]{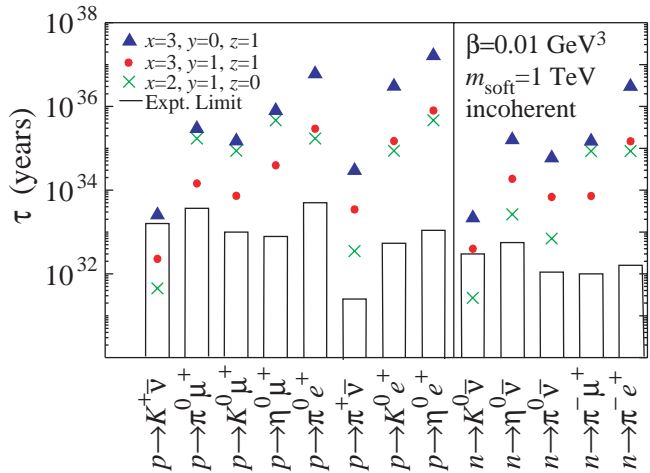}
  \caption{Plot of nucleon lifetime in years for eight proton 
    decay modes (left side) and five neutron decay modes (right side).
    The different symbols represent different
    $U(1)_X$ charge assignments, namely Model~1
  ({\color{Red}$\bullet$}) which has
    $x=3,\ y=-1,\ z=1$, Model~2 ({\color{Green}$\times$}) with $x=2,\
    y=1,\ z=0$, and Model~3 ({\color{Blue}$\blacktriangle$}) with $x=3,\
    y=0,\ z=1$. The experimental limit for each mode is shown as a
    vertical column.}
  \label{fig:3models}
\end{figure}

To further illustrate the differences in the pattern of decay modes
from different models, we will focus on three specific models, namely
Model~1 with $x=3,\ y=-1,\ z=1$, Model~2 with $x=2,\ y=1,\ z=0$, and
Model~3 with $x=3,\ y=0,\ z=1$. Model~3 was specifically studied
in~\cite{Dreiner:2003yr}. Model 2 is has a particularly nice charge
assignment as shown in Table~\ref{TableII}.  In
Figure~\ref{fig:3models} we show the normalized partial lifetimes for
these three models in eight proton decay modes (left side) and five
neutron decay modes (right side).

Various modes, if discovered, may serve as a discriminator between
models. For example, among the three models shown in
Figure~\ref{fig:3models}, any mode involving a muon in the final state
can differentiate Model 1 from Models 2 and 3. The decay mode $p\to
\pi^0 e^+$ can differentiate Model 2 from Model 3, etc.

\section{Conclusion}
\label{sec:conclusion}

\noindent
We have shown that the search for nucleon decay is a powerful probe of
physics at the Planck scale. We focused on a a class of ambitious,
string motivated, Froggatt-Nielsen models. These models explain the
masses and mixings of all SM fermions while automatically enforcing
$R$-parity as an accidental symmetry.  In the context of these models,
we have shown that operators suppressed by the Planck scale lead to
nucleon decay rates that are generically right near the current
experimental limits, even without Grand Unification. In fact, current
bounds disfavor many of the 24 distinct models of this type.

The main unknown in this program is whether supersymmetry exists, and
if so, what the mass spectrum is.  Data from the LHC and a Linear
Collider, along with improved lattice calculations of the matrix
element, will remove much of the uncertainty (see
Fig. \ref{fig:Knu-uncertainties}) leaving us with much tighter
constraints.  Furthermore, discovery of several proton decay modes
would serve as a good discriminator between the various model
parameters. Similarly, a measurement of $\theta_{13}$ in neutrino
experiments, or of $\tan \beta$ at a collider, can help narrowing down
the choice of viable models.  We thus conclude that upcoming
experiments may truly be probing physics at the Planck scale.

\noindent
\acknowledgments{We thank R.~N.~Mohapatra for pointing out
  Ref.~\cite{Kearns}. RH and DTL thank the Institute for Advanced
  Study for
  hospitality. MT greatly appreciates that his work was supported by a
  fellowship within the Postdoc-Programme of the German Academic
  Exchange Service (Deutscher Akademischer Austauschdienst, DAAD).
  The work of HM was supported in part by the Institute for Advanced
  Study, funds for Natural Sciences.  DTL, RH and HM were supported in
  part by the DOE under contract DE-AC03-76SF00098 and in part by NSF
  grant PHY-0098840.}

\appendix

\section{Dressing Diagrams}
\label{sec:dressing}

\noindent
The superpotential operators \QQQL and \UUDE are not all independent.
In particular, 
\begin{eqnarray}
\qqql{i}{j}{k}{l}&=&\qqql{j}{i}{k}{l} \nonumber \\
\qqql{i}{j}{i}{l}&=&-\frac12 \qqql{i}{i}{j}{l}\ \mbox{ (no sum) } \\
\uude{i}{j}{k}{l}&=&-\uude{j}{i}{k}{l}. \nonumber
\end{eqnarray}
Furthermore, contributions to
nucleon decay can only come from operators with at least one first
generation superfield. The list of independent operators that
contribute to proton decay is shown in Table~\ref{tab:operators}.
Since we do not know the exact \orderone coefficients in front of
these operators, we will simply work with an independent basis of
operators and assume that their unknown coefficients are given by the
$\eps$-suppression determined by the $X$-charges times \orderone
numerical factors, excluding accidental cancellations.
\begin{table}[ht!]
  \begin{tabular}{|l|l|}
    \hline
    Operator & Independent set of $(ijkl)$ \\ \hline
    $Q_i Q_j Q_k L_l$ & $(112l), (113l), (221l), (331l), (123l), (132l),
    (231l)$ \\
    $\bbar{U}_i \bbar{U}_j \bbar{D}_k \bbar{E}_l$ & $(121l), (122l), 
    (123l),
    (131l), (132l), (133l)$\\ \hline
  \end{tabular}
  \caption{Independent operators relevant for nucleon decay as labeled
    by their generation indices.} 
\label{tab:operators}
\end{table}
The operators are then dressed by a gaugino or higgsino as shown in
Figure~\ref{fig:dress}. In general, there are many ways each diagram
can be dressed. However, it has been long known that in the limit of
degenerate squarks many dressing combinations cancel due to a Fierz
identity~\cite{Belyaev:ik,Babu:1995cw}. Here we demonstrate this
cancellation explicitly using gluino dressing as an example, following the
description in~\cite{Goh:2003nv}. Because gluinos are flavor blind,
they couple to up- and down-type squarks equally and do not change
flavor.  Since they do not couple to leptons, there are three ways
that gluinos can dress a given operator $\bbar{U}_i \bbar{U}_j
\bbar{D}_k \bbar{E}_l$.
\begin{eqnarray} \label{eqn:gluinoops} 
  1: & & f(\bbar{U}_i,\bbar{U}_j)\; (\bbar{E}_l
  \bbar{D}_k)(\bbar{U}_i \bbar{U}_j) \nonumber \\ 
  2: & & f(\bbar{U}_i,\bbar{D}_k)\; (\bbar{E}_l \bbar{U}_j)(\bbar{D}_k
  \bbar{U}_i) \\
  3: & & f(\bbar{D}_k,\bbar{U}_j)\; (\bbar{E}_l
  \bbar{U}_i)(\bbar{U}_j \bbar{D}_K) \nonumber 
\end{eqnarray}
In these expressions the fields are Weyl spinors contracted within the
parentheses. The function $f(x,y)$ comes from the loop integral and
depends on the masses of the gluino and the squarks $x$ and $y$. Thus
$f(x,y)$ encodes all the flavor dependence, so in the limit of
degenerate squarks each factor of $f$ above is common to all three
terms. The
interesting fact is that the sum of those three operators vanishes by
the following Fierz identity: \beq (AB)(CD) + (AD)(BC) + (AC)(DB) = 0,
\eeq which can be easily shown by rewriting the epsilon tensors used
to contract the pairs of Weyl spinors in terms of Kronecker delta
functions. Thus the sum of the three
operators in Eqn.~(\ref{eqn:gluinoops}) will vanish whenever they all
have the same coefficient out front, which occurs in the degenerate
squark limit. This can also be explained in words by noting that in
the final four-fermion operator the three quark fields must be
antisymmetric in color, antisymmetric in flavor (not generation, but
\emph{flavor}) and, since they are fermions, antisymmetric in spin as
well to make the total operator antisymmetric under fermion exchange.
But since there are three quark fields but only two spin states for each
Weyl spinor, there is no way to make an operator that is completely
antisymmetric in spin, hence the whole collection must vanish.

This argument holds for all the gluino dressings of \QQQL and
$\bbar{U}_i\bbar{U}_j\bbar{D}_k\bbar{E}_l$. In addition, it also holds
for bino dressing of $Q_iQ_jQ_kL_l$~\cite{Goh:2003nv}. The bino and
neutral higgsino
dressings of \UUDE do not lead to any operators that contribute to
nucleon decay, due to the inevitable presence of 2nd or 3rd generation
up-quarks. Thus the only relevant dressing of the $SU(2)$ singlet
fields in \UUDE is by charged higgsinos. The \QQQL operators get
contributions from charged and neutral winos and higgsinos. In
general, the dominant contributions are usually from wino dressing of
\QQQL, except for the \UUDE contribution to the
$K\nu_\tau$ final state when $\tan\beta$ is large, as pointed out
by~\cite{Goto:1998qg}.

\section{TRANSFORMING INTO THE MASS BASIS}
\label{sec:kahler}

\noindent
Once the flavon acquires a VEV the coefficients $C^{ijkl}_{L,R}$ are
each determined by a dimensionless $\mathcal{O}(1)$ coefficient times
an $\eps$-suppression. In principle, some coefficients may be exactly
zero due to negative overall $X$-charge, the so-called
``supersymmetric zeros'', but this does not occur in the models we
consider. The naive $C^{ijkl}_{L,R}$ cannot be
directly plugged into Eqn.~(\ref{eqn:decaywidth}). First one needs to
take into account two superfield-transformations, namely the
canonicalization of the K\"ahler potential (CK), and then the
transformation into the mass basis of the quarks and leptons (TM):
\begin{enumerate}
\item From the outset the K\"ahler potential need not have the
  canonical form (see
  \cite{Leurer:1992wg,Feinberg:1959ui,Binetruy:1996xk,
    Dreiner:2003hw,Jack:2003pb} for details). For example, the K\"ahler
potential for the quark doublets takes the form 
  \begin{equation}
    \overline{Q}_i~{H_{\!Q}}^{\!ij}~Q_j~=~
    \overline{\bsym{\Big[{\bsym{\mathcal{C}_{\!Q}}}\cdot Q}\Big]}_i 
    ~~ \delta^{ij} ~~
    \Big[ \bsym{{\bsym{\mathcal{C}_{\!Q}}}\cdot Q}  \Big]_j
  \end{equation}
  where $\bsym{H_{\!Q}}$ is a Hermitian matrix with hierarchical
  entries, generated when the flavon acquires a VEV. It can be
  diagonalized by the matrix $\bsym{\mathcal{C}_{\!Q}}$. This
  redefinition, \emph{i.e.} the ``CK'', affects the coupling constants of the
  superpotential, \emph{e.g.}
  \begin{eqnarray}
    \bsym{G_{\!U}}~\rightarrow~\bsym{{G_{\!U}}^{\!C\!K}}=
    \frac{1}{\sqrt{ H^{\left(H^\mcal{U}
          \right)}~}}~{{\bsym{\mathcal{C}_{\!Q}}}^{-1}}^T\cdot\bsym{G_{\!U}}
    \cdot{\bsym{\mathcal{C}_{\!\overline{U}}}}^{-1}.
  \end{eqnarray}
\item Transforming the superfields to the mass basis, the coupling
constants are then subject to the ``TM''.  One has \beq
  \boldsymbol{{G_{\!U}}^{\!C\!K}}~=~
  \boldsymbol{\mathcal{U}_{U_{\!L}}}^{\!T}\cdot\mbox{{\bf{diag}}}
  \{m_u,m_c,m_t\}\cdot\boldsymbol{\mathcal{U}_{\overline{U_{\!R}}}},
  \eeq etc., the $\boldsymbol{\mathcal{U}}_{...}$ being unitary.
\end{enumerate}
The other superpotential coupling constants have to be transformed
correspondingly, \emph{e.g.}
\begin{widetext}
\begin{eqnarray}
  C^{ijkl}_{R}\rightarrow {C_R^{C\!K\!+\!T\!M}}^{~ijkl}=
  \Big[\big(\boldsymbol{\mathcal{U}_{\overline{U_{\!R}}}}
  \cdot   \bsym{\mathcal{C}_{\!\overline{U}}}   \big)^{-1}
  \Big]^{~i}_{\!i^\prime}~\Big[\big(\boldsymbol{
    \mathcal{U}_{\overline{U_{\!R}}}}\cdot   \bsym{
    \mathcal{C}_{\!\overline{U}}}   \big)^{-1}\Big]^{~j}_{\!j^\prime}~
  \Big[\big(\boldsymbol{\mathcal{U}_{\overline{D_{\!R}}}}\cdot   
  \bsym{\mathcal{C}_{\!\overline{D}}}   \big)^{-1}\Big]^{~k}_{
    \!k^\prime}~\Big[\big(\boldsymbol{\mathcal{U}_{\overline{E_{\!R}}}}
  \cdot   \bsym{\mathcal{C}_{\!\overline{E}}}   \big)^{-1}
  \Big]^{~l}_{\!l^\prime}~C^{i^\prime j^\prime k^\prime l^\prime }_{R}.
\end{eqnarray}
\end{widetext}

As we lack knowledge of the $\mathcal{O}(1)$ coefficients, this can
only be done approximately, supposing no accidental cancellations. One
can show \cite{Binetruy:1996xk} that
\begin{equation}\label{eq:ck}
  \Big[\bsym{{{\mathcal{C}_{\!Q}}}}^{-1}\Big]_{\!ij}~\sim~
  \eps^{|X_{Q^i}-X_{Q^j}|}, ~~\mbox{etc.},
\end{equation}
from which it follows that the CK fills up supersymmetric zeros
but does not change the $\eps$-suppression of any other nonzero
entries. Using Table~\ref{tab:xcharges} this gives 
\begin{eqnarray}  
  \bsym{{{\mathcal{C}_{\!Q}}}}^{-1}&\sim&
  \left(\begin{array}{ccc}
      1 &\eps^{|1+y|} &\eps^{|3+y|} \\
      \eps^{|1+y|} & 1 &\eps^{2} \\
      \eps^{|3+y|} & \eps^{2} &1
    \end{array}\right),
\end{eqnarray}
\begin{eqnarray}
  \bsym{{{\mathcal{C}_{\!L}}}}^{-1}&\sim&\left(\begin{array}{ccc}
      1 &\eps^{z} &\eps^{z} \\
      \eps^{z} & 1 & 1 \\
      \eps^{z} & 1 &1
    \end{array}\right),
\end{eqnarray}
\begin{eqnarray}
  \bsym{{{\mathcal{C}_{\!\overline{D}}}}}^{-1} &\sim&
  \left(\begin{array}{ccc}
      1 &\eps^{|1-y|} &\eps^{|1-y|} \\
      \eps^{|1-y|} & 1 & 1 \\
      \eps^{|1-y|} & 1 &1
    \end{array}\right),
\end{eqnarray}
\begin{eqnarray}
  \bsym{{{\mathcal{C}_{\!\overline{U}}}}}^{-1}&\sim&
  \left(\begin{array}{ccc}
      1 &\eps^{|3-y|} &\eps^{|5-y|} \\
      \eps^{|3-y|} & 1 &\eps^{2} \\
      \eps^{|5-y|} & \eps^{2} &1
    \end{array}\right),
\end{eqnarray}
\begin{eqnarray}
  \bsym{{{\mathcal{C}_{\!\overline{E}}}}}^{-1}  &\sim&
  \left(\begin{array}{ccc}
      1 &\eps^{2} &\eps^{4} \\
      \eps^{2} & 1 &\eps^{2} \\
      \eps^{4} & \eps^{2} &1
    \end{array}\right).
\end{eqnarray}
The $\boldsymbol{\mathcal{U}}_{...}$ are given in
Table~\ref{tab:mass-basis}, having used the expressions in
\cite{Hall:1993ni}.
\begin{table*}[ht!]
  \begin{tabular}{|c|c|c|c|c|}
    \hline
    $\boldsymbol{\mathcal{U}_{U_{\!L}}}^*,~\boldsymbol{
      \mathcal{U}_{D_{\!L}}}^*$ &
    $\boldsymbol{\mathcal{U}_{\overline{U_{\!R}}}}^\dagger$
    & $\boldsymbol{\mathcal{U}_{\overline{D_{\!R}}}}^\dagger$ & 
    $\boldsymbol{\mathcal{U}_{E_{\!L}}}^*,~\widetilde{
      \boldsymbol{\mathcal{U}_{N_{\!L}}}}^* ~$ & $\boldsymbol{
      \mathcal{U}_{\overline{E_{\!R}}}}^\dagger$   \\
    \hline
    $~\left(\begin{array}{ccc}
        1 &\eps^{1+y} &\eps^{3+y}\\
        \eps^{1+y} & 1 &\eps^{2} \\
        \eps^{3+y} & \eps^{2} &1
      \end{array}\right)^{~\phantom{\Big|}}_{~\phantom{\Big|}}$ &
    $~\left(\begin{array}{ccc}
        1 &\eps^{3-y} &\eps^{5-y} \\
        \eps^{3-y} & 1 &\eps^{2} \\
        \eps^{5-y} & \eps^{2} &1
      \end{array}\right) ~$ & $~\left(\begin{array}{ccc}
        1 &\eps^{1-y} &\eps^{1-y} \\
        \eps^{1-y} & 1 & 1 \\
        \eps^{1-y} & 1 &1
      \end{array}\right) ~$ & $~\left(\begin{array}{ccc}
        1 &\eps^{z} &\eps^{z} \\
        \eps^{z} & 1 & 1 \\
        \eps^{z} &  1 &1
      \end{array}\right)^{~\phantom{\Big|}}_{~\phantom{\Big|}}$  &
    $~\left(\begin{array}{ccc}
        1 &\eps^{2} &\eps^{4} \\
        \eps^{2} & 1 &\eps^{2} \\
        \eps^{4} & \eps^{2} &1
      \end{array}\right)^{~\phantom{\Big|}}_{~\phantom{\Big|}} $ \\
    \hline
  \end{tabular}
  \caption{\label{tab:mass-basis}  
    The matrices needed to perform the TM.  For the quarks they are
    valid for any $X$-charge assignment which leads to the five pairs
    of quark mass matrices displayed in \cite{Dreiner:2003yr}. The
    ones for the leptons ($y$-independent)  are only valid for the
    $X$-charge assignment used in  this article.
    $\widetilde{\boldsymbol{\mathcal{U}_{N_{\!L}}}}^*$ is the almost
    unitary matrix  which Schur-diagonalizes the symmetric mass matrix
    of the light neutrinos. Note that these approximate TM matrices
    reproduce  the CKM matrix and the MNS matrix, as they should.  } 
\end{table*}

Since the $C_{R,L}^{ijkl}$ we consider do not have any supersymmetric
zeros, their $\eps$-dependence is not changed by the CK. As for TM,
comparing Table~\ref{tab:mass-basis} with the $\boldsymbol{
  \mathcal{C}_{...}}$ given above, one finds \emph{e.g.}
$~\boldsymbol{\mathcal{U}_{U_{\!L}}}^*
\sim\boldsymbol{\mathcal{C}_{\!Q}}~$ and
$~\boldsymbol{\mathcal{U}_{\overline{U_{\!R}}}}^\dagger
\sim\boldsymbol{\mathcal{C}_{{\!\overline{U}}}}$.  Since
$\boldsymbol{\mathcal{C}_{\!Q}}^2\sim \boldsymbol{\mathcal{C}_{\!Q}}$
(and likewise for the others), we find that the $C^{ijkl}_{L,R}$ are
not changed substantially when going into the mass basis either.
\emph{Thus the naively calculated $\epsilon$-suppression for
  $C^{ijkl}_{L,R}$
  remains the same after changing to the mass basis.}\\



\begin{thebibliography}{99}

\bibitem{Sakharov:dj}
A.~D.~Sakharov,
Pisma Zh.\ Eksp.\ Teor.\ Fiz.\  {\bf 5}, 32 (1967)
[JETP Lett.\  {\bf 5}, 24 (1967\ SOPUA,34,392-393.1991\ 
UFNAA,161,61-64.1991)].

\bibitem{Kearns}
E.~Kearns, Talk at Snowmass 2001,\\
\mbox{http://hep.bu.edu/\~{}kearns/pub/kearns-pdk-snowmass.pdf}.

\bibitem{Murayama:2001ur}
H.~Murayama and A.~Pierce,
Phys.\ Rev.\ D {\bf 65}, 055009 (2002)
[arXiv:hep-ph/0108104].

\bibitem{Murayama:1994tc}
H.~Murayama and D.~B.~Kaplan,
Phys.\ Lett.\ B {\bf 336}, 221 (1994)
[arXiv:hep-ph/9406423].

\bibitem{Froggatt:1978nt}
C.~D.~Froggatt and H.~B.~Nielsen,
Nucl.\ Phys.\ B {\bf 147}, 277 (1979).

\bibitem{Dreiner:2003hw}
H.~K.~Dreiner and M.~Thormeier,
Phys.\ Rev.\ D {\bf 69}, 053002 (2004)
[arXiv:hep-ph/0305270].

\bibitem{Kakizaki:2002hs}
M.~Kakizaki and M.~Yamaguchi,
JHEP {\bf 0206}, 032 (2002)
[arXiv:hep-ph/0203192].

\bibitem{Arkani-Hamed:1999dc}
N.~Arkani-Hamed and M.~Schmaltz,
Phys.\ Rev.\ D {\bf 61}, 033005 (2000)
[arXiv:hep-ph/9903417];

\bibitem{Kakizaki:2001ue}
M.~Kakizaki and M.~Yamaguchi,
arXiv:hep-ph/0110266.

\bibitem{Dreiner:2003yr}
H.~K.~Dreiner, H.~Murayama and M.~Thormeier,
arXiv:hep-ph/0312012.

\bibitem{Green:sg}
M.~B.~Green and J.~H.~Schwarz,
Phys.\ Lett.\ B {\bf 149}, 117 (1984).

\bibitem{Dine:bq}
M.~Dine, N.~Seiberg, X.~G.~Wen and E.~Witten,
Nucl.\ Phys.\ B {\bf 289}, 319 (1987).

\bibitem{Dine:1986zy}
M.~Dine, N.~Seiberg, X.~G.~Wen and E.~Witten,
Nucl.\ Phys.\ B {\bf 278}, 769 (1986).

\bibitem{Atick:1987gy}
J.~J.~Atick, L.~J.~Dixon and A.~Sen,
Nucl.\ Phys.\ B {\bf 292}, 109 (1987).

\bibitem{Dine:1987gj}
M.~Dine, I.~Ichinose and N.~Seiberg,
Nucl.\ Phys.\ B {\bf 293}, 253 (1987).

\bibitem{Espinosa:2004ya}
J.~R.~Espinosa and A.~Ibarra,
arXiv:hep-ph/0405095.

\bibitem{King:2004tx}
S.~F.~King, I.~N.~R.~Peddie, G.~G.~Ross, L.~Velasco-Sevilla and O.~Vives,
arXiv:hep-ph/0407012.

\bibitem{Hall:1999sn}
L.~J.~Hall, H.~Murayama and N.~Weiner,
Phys.\ Rev.\ Lett.\  {\bf 84}, 2572 (2000)
[arXiv:hep-ph/9911341],

\bibitem{Haba:2000be}
N.~Haba and H.~Murayama,
Phys.\ Rev.\ D {\bf 63}, 053010 (2001)
[arXiv:hep-ph/0009174].

\bibitem{Sato:1997hv}
J.~Sato and T.~Yanagida,
Phys.\ Lett.\ B {\bf 430}, 127 (1998)
[arXiv:hep-ph/9710516].

\bibitem{Giudice:1988yz}
G.~F.~Giudice and A.~Masiero,
Phys.\ Lett.\ B {\bf 206}, 480 (1988).

\bibitem{Kim:1994eu}
J.~E.~Kim and H.~P.~Nilles,
Mod.\ Phys.\ Lett.\ A {\bf 9}, 3575 (1994)
[arXiv:hep-ph/9406296].

\bibitem{Goto:1998qg}
T.~Goto and T.~Nihei,
Phys.\ Rev.\ D {\bf 59}, 115009 (1999)
[arXiv:hep-ph/9808255].

\bibitem{Claudson:1981gh}
M.~Claudson, M.~B.~Wise and L.~J.~Hall,
Nucl.\ Phys.\ B {\bf 195}, 297 (1982).

\bibitem{Buras:1977yy}
A.~J.~Buras, J.~R.~Ellis, M.~K.~Gaillard and D.~V.~Nanopoulos,
Nucl.\ Phys.\ B {\bf 135}, 66 (1978).

\bibitem{Kuramashi:2000hw}
Y.~Kuramashi  [JLQCD Collaboration],
arXiv:hep-ph/0103264.

\bibitem{Aoki:2002ji}
Y.~Aoki  [RBC Collaboration],
Nucl.\ Phys.\ Proc.\ Suppl.\  {\bf 119}, 380 (2003)
[arXiv:hep-lat/0210008].

\bibitem{Hagiwara:fs}
K.~Hagiwara {\it et al.}  [Particle Data Group Collaboration],
Phys.\ Rev.\ D {\bf 66}, 010001 (2002).

\bibitem{Babu:1995cw}
K.~S.~Babu and S.~M.~Barr,
Phys.\ Lett.\ B {\bf 381}, 137 (1996)
[arXiv:hep-ph/9506261].

\bibitem{Belyaev:ik}
V.~M.~Belyaev and M.~I.~Vysotsky,
Phys.\ Lett.\ B {\bf 127}, 215 (1983).

\bibitem{Goh:2003nv}
H.~S.~Goh, R.~N.~Mohapatra, S.~Nasri and S.~P.~Ng,
arXiv:hep-ph/0311330.

\bibitem{Leurer:1992wg}
M.~Leurer, Y.~Nir and N.~Seiberg,
Nucl.\ Phys.\ B {\bf 398}, 319 (1993)
[arXiv:hep-ph/9212278].

\bibitem{Feinberg:1959ui}
G.~Feinberg, P.~Kabir and S.~Weinberg,
Phys.\ Rev.\ Lett.\  {\bf 3}, 527 (1959).

\bibitem{Binetruy:1996xk}
P.~Bin\'{e}truy, S.~Lavignac and P.~Ramond,
Nucl.\ Phys.\ B {\bf 477}, 353 (1996)
[arXiv:hep-ph/9601243].

\bibitem{Jack:2003pb}
I.~Jack, D.~R.~T.~Jones and R.~Wild,
Phys.\ Lett.\ B {\bf 580}, 72 (2004)
[arXiv:hep-ph/0309165].

\bibitem{Hall:1993ni}
L.~J.~Hall and A.~Rasin,
Phys.\ Lett.\ B {\bf 315}, 164 (1993)
[arXiv:hep-ph/9303303].

\end{thebibliography}
\end{document}